\documentclass[
aps,nofootinbib,
showpacs,showkeys,preprint
tightenlines,preprintnumbers,] {revtex4}

\usepackage{epsf,epsfig,subfigure,graphicx,amsmath,amssymb 
}
\usepackage{color}
\newcommand{\dis}[1]{\begin{equation}\begin{split}#1\end{split}\end{equation}}
\newcommand{\ie}{{\it i.e.~}}
\newcommand{\etal}{{\it et al.\,}}

\newcommand{\Qem}{Q_{\rm em}}

\newcommand{\Uanom}{U(1)$_{\rm anom}$}

\def\sw0{{$\sin^2\theta_W^0$}}

\def\dell{\delta_{\rm PMNS}}
\def\delks{\delta_{\rm KS}}

\def\Vckm{V_{\rm CKM}}
\def\Vpmns{V_{\rm PMNS}}

\newcommand{\Z}{{\bf Z}}

\def\E6{{\rm E_6}}

\def\EE8{{\rm E_8\times E_8'}}

\def\flip{SU$(5)_{\rm flip}$}

\def\three{{\bf 3}}

\def\one{{\bf 1}}

\def\five{{\bf 5}}
\def\ten{{\bf 10}}
\def\tenb{{\overline{\bf 10}}}

\def\fiveb{{\overline{\bf 5}}}

\def\three{{\bf 3}}

\begin{document}

\draft

\title{\Large\bf Leptonic CP violation in flipped SU(5)  GUT from $\Z_{12-I}$ Orbifold Compactification}

\author{Junu Jeong$^{(1)}$, Jihn E.  Kim$^{(1,2,3)}$,  Soonkeon Nam$^{(2)}$}

\address
{$^{(1)}$Center for Axion and Precision Physics Research (Institute of Basic Science), KAIST Munji Campus, 193 Munjiro,  Daejeon 34051, Republic of Korea,   \\
$^{(2)}$Department of Physics, Kyung Hee University, 26 Gyungheedaero, Dongdaemun-Gu, Seoul 02447, Republic of Korea,  \\
 $^{(3)}$ Department of Physics and Astronomy, Seoul National University, 1 Gwanakro, Gwanak-Gu, Seoul 08826, Republic of Korea
}

\begin{abstract} 
We obtain a phenomenologically acceptable PMNS matrix in a flipped SU(5) model inspired by the  compactification of heterotic string $\EE8$. To analyze the Jarlskog determinant efficiently, we include the simple Kim-Seo form for the   Pontecorbo-Maki-Nakagawa-Sakata matrix.  We also noted that  $|\dell|\lesssim 64^{\rm o}$ for the normal hierarchy of neutrino masses with the PDG book parametrization.
 
\end{abstract}
\maketitle

\section{Introduction}\label{sec:Introduction}
 
The most urgent theoretical issue in the standard model(SM) is probing the symmetry structure from which  the observed flavor phenomena can be understood.  It is desirable if such symmetry results from an ultra-violet completed theory such as from string compactification \cite{Candelas,Dixon2,Ibanez1,Tye87,Bachas87,Gepner87}. Most studies along this direction were centered on obtaining three families \cite{IKNQ,Munoz88,Ellis89,KimKyae07}.\footnote{For more  references, see Ref. \cite{Kim18Rp}.}
Now time is ripe enough to study the details of the flavor structure from string compactification. In the quark sector,  the Cabibbo-Kobayashi-Maskawa matrix \cite{Cabibbo63, KM73} has been studied in our previous paper \cite{Junu18}.

In this paper, we present a numerical study on  the Pontecorbo-Maki-Nakagawa-Sakata(PMNS) matrix \cite{PMNS1,PMNS2} from string compactification via many U(1)'s arising in string compactification. String compactification in our example allows all the needed Yukawa couplings in the SM as non-renormalizable ones  \cite{Kim18Rp}. Therefore, the grand design \cite{Munoz12} of relating the neutrino masses with the magnitude of the $\mu$ term with renormalizable terms only is not applicable here. The relation might be intertwined in an elaborate way in our model since the $\Z_{4R}$ discrete symmetry automatically gives the $\mu$ term at the electroweak scale \cite{Lee11}. Also, we are far from obtaining  non-Abelian discrete symmetries such as $A_4$ \cite{Ma03} in this study.

The first step in modeling a flavor structure from string compactification is to allocate the CP phase at some convenient slots in the mass matrices of neutrinos and charged leptons. Toward this, we point out that it is useful to use the Kim-Seo(KS) parametrization \cite{KimSeo11} of the CKM and PMNS matrices. With the KS parametrization, the CP phase in the Jarlskog triangle can be put in the (31) elements of the CKM and PMNS matrices $V^{\rm KS}$, and the physical magnitude of CP violation is looked upon just from this simple matrix because the Jarlskog determinant is $J=-{\rm Im\,}V^{\rm KS}_{31}V^{\rm KS}_{22}V^{\rm KS}_{13}$ \cite{KimSeo12}. Toward a model building, the next step is to obtain phenomenologically acceptable mass matrices. Since the Yukawa couplings are too many in standard-like models, here we work in the flipped SU(5) GUT \cite{Barr82,DKN84} where the number of Yukawa couplings are much less than in the standard-like models. In the flipped SU(5), the neutrino mass matrix turns out to be symmetric and hence we propose in this paper to put the CP phase in the charged lepton mass matrix.

In Sec. \ref{sec:JarlskogT}, we recapitulate the simple form of the Jarlskog determinant.  In Sec. \ref{sec:Mdiagonal}, we obtain a useful fit of the PMNS matrix in the KS form from the data presented in the Particle Data Book \cite{PDG18pmns}. As a by-product, we will observe that $|\dell|<62.8^{\rm o}$ for the normal hierarchy of neutrino masses with the PDG book parametrization \cite{PDG18pmns}. In Sec. \ref{sec:FSU5}, we present possible terms of neutrino and charged lepton mass terms allowed by the quantum numbers of Ref. \cite{Kim18Rp}.   
Then, we locate possible phases in the complex vacuum expectation values(VEVs) of the SM singlet fields $\sigma_i$. Section \ref{sec:Conclusion} is a brief conclusion.

\section{Jarlskog determinant}\label{sec:JarlskogT}

\begin{figure}[!t]
\includegraphics[width=0.35\textwidth]{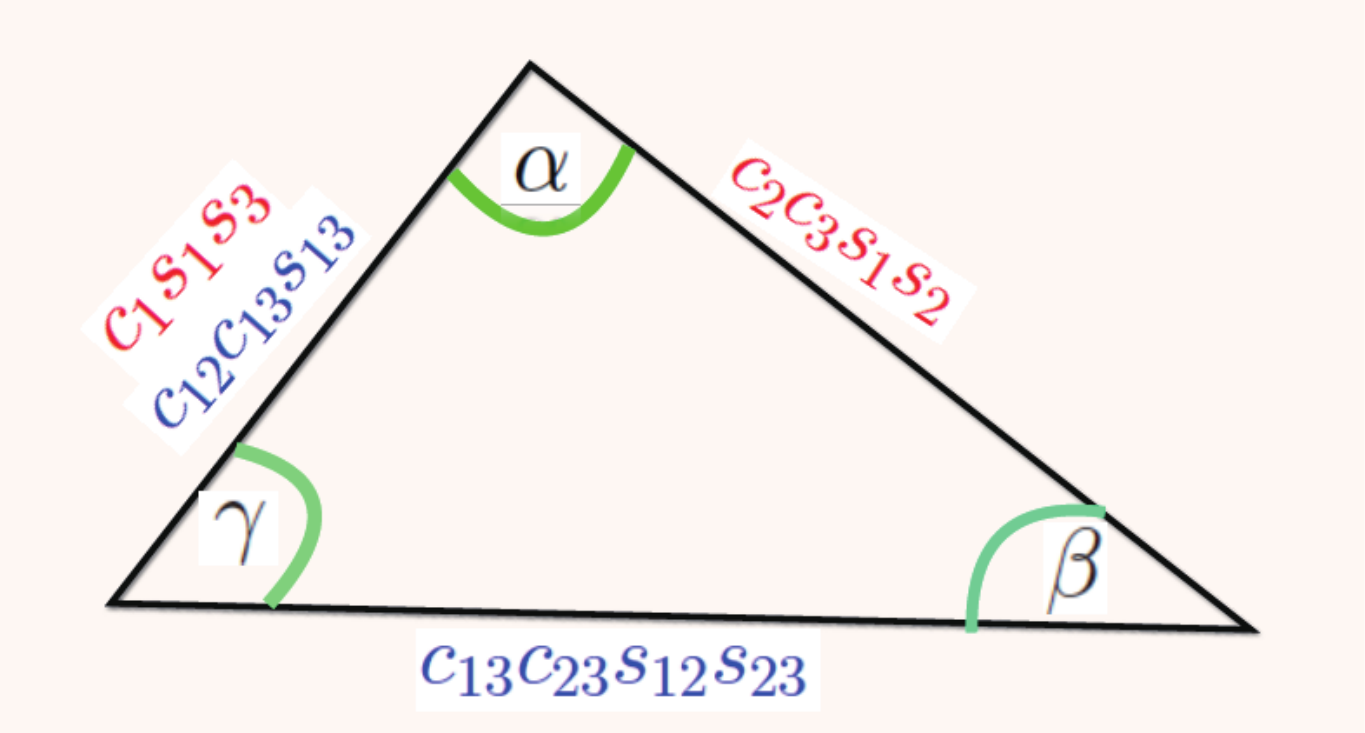} 
 \caption{The Jarlskog triangle \cite{Jarlskog85,KimCorfu16} where $c_i=\cos\theta_i, s_i=\sin\theta_i, $ and $c_{ij}=\cos\theta_{ij},s_{ij}=\sin\theta_{ij} $.} \label{fig:Jarlskog} 
\end{figure}

There are four angles, three real angles and one phase, describing charged current (CC) weak interactions both in the quark sector \cite{KM73} and in the lepton sector \cite{PMNS1,PMNS2}. The importance of the CP violation is given by the so-called Jarlskog determinant $J$ which is twice the area of the triangle shown in Fig. \ref{fig:Jarlskog}.
Originally, it was derived by considering specific modes of weak processes, for example $\sum_i V^*_{i2}V_{i3}=0$ which defines a unitary triangle, and the initial estimate of $J$ was given by a sum of four products of $V_{ij}$ in the form $V^* V V^* V$ \cite{Jarlskog85} where each $V^*V$ has a process significance. Therefore, we can consider six triangles, three from two columns and three from two rows.

But, it is known that by making Det.$\,V$ real, one can express $J=-{\rm Im\,}V^{\rm KS}_{31}V^{\rm KS}_{22}V^{\rm KS}_{13}$ or $ J={\rm Im\,}V^{\rm KS}_{11}V^{\rm KS}_{22}V^{\rm KS}_{33}$, etc. \cite{KimSeo12}. In this form, $J$ relates the entire range of the $3\times 3$ matrix and hence it can be a theory dependent number. So, it is better to use this form of $J$.   Figure \ref{fig:Jarlskog} is drawn by considering $V^*_{i1}V_{i3}$.
In the particle data book, the CKM matrix is defined by the $W^+_\mu$ coupling \cite{PDG18ckm} while  the PMNS matrix is defined by the $W^-_\mu$ coupling   \cite{PDG18pmns,Valle80}.  
Three real angles used in the PDG book are $\theta_{12},\theta_{13}$, and $\theta_{23}$ \cite{PDG18ckm} and the phase is denoted as $\delta$. 

The angles $\alpha, \beta,$ and $\gamma$ of  Fig. \ref{fig:Jarlskog} are related to $\delta$ of $V$. The same area of the triangle can be given in a different parametrization also. One useful parametrization, the Kim-Seo (KS)  parametrization with Det.$\,V^{\rm KS}=1$, locates three real numbers in the first row  \cite{KimSeo11}, 

\dis{
V_{\rm KS}= \left(\begin{array}{ccc} {\color{red}R_1},&{\color{red}R_2},&{\color{red}R_3}  \\ [0.2em]
  T_1, &R_4+R_5e^{-i\delta},&
T_2 \\[0.2em]
 R_6e^{i\delta}  ,& T_3 ,& T_4
\end{array}\right),\label{eq:KSq}
}
where $R_i$  and $T_i$ are  real and  complex numbers, respectively. Then, we have  $J=- R_3R_4R_6\sin\delta$.
It is remarkable to note that   real numbers for  one row (shown with the red color) makes it possible  to visualize the importance of $e^{i\delta}$ in the position   $V_{31}$  \cite{KimSeo12}. A complex (22) element can be always written in the form separating out the term with the factor $e^{-i\delta}$.   Therefore, to use the analyses in the PDG book with the simple form given in Eq. (\ref{eq:KSq}), we solve, in view of Fig. \ref{fig:Jarlskog},  the equations for $\theta_i$ in terms of $\theta_{ij}$,
\dis{
&c_1s_1s_3= c_{12}c_{13}s_{13},\\
& c_2^2c_3^2s_1^2s_2^2
 + c_1^2s_1^2s_3^2-2c_1c_2c_3s_1^2s_2s_3\cos\alpha =c_{13}^2c_{23}^2s_{12}^2s_{23}^2,\\  
&c_2^2c_3^2s_1^2s_2^2=c_{13}^2c_{23}^2s_{12}^2s_{23}^2
 +c_{12}^2c_{13}^2s_{13}^2- 2c_{12}c_{13}^2c_{23} s_{12}
 s_{23} s_{13}\cos\gamma .\label{eq:angles}
}
For the fixed triangle given by (\ref{eq:angles}), the area relation results in
\dis{
c_2c_3s_1s_2\sin\alpha= c_{13}c_{23}s_{12}s_{23} \sin\gamma.\label{eq:arearelation}
}  
Since there are four parameters to be determined \ie $\theta_{1,2,3}$ and $\alpha$ from  Eq. (\ref{eq:angles}), there is a degree of freedom to define the KS form from the observed angles in the PDG book. Even if we can determine the KS parameters from (\ref{eq:angles}) with one degeneracy  parameter, the additional relation (\ref{eq:arearelation}) has a profound meaning. It must be satisfied for all real values of parameters $\theta_{i}, \alpha$ and  $\theta_{ij}, \gamma$. For some angles, therefore, there must be a bound for the relation  (\ref{eq:arearelation}) to be satisfied.
 Let us fix the parametrization such that the (11) element in the KS form agrees with the (11) element of the PDG book, $c_1=c_{12}c_{13}$.
Then, the four conditions to determine the KS parameters are 
\dis{
&c_1=c_{12}c_{13},\\
 &s_1s_3=s_{13},\\
 &\sqrt{c_2^2c_3^2s_1^2s_2^2
 + c_1^2s_1^2s_3^2-2c_1c_2c_3s_1^2s_2s_3\cos\alpha} =\pm c_{13}c_{23}s_{12}s_{23},\\
 &c_2c_3s_1s_2=\pm\sqrt{c_{13}^2c_{23}^2s_{12}^2s_{23}^2 +c_{12}^2c_{13}^2s_{13}^2- 2c_{12}c_{13}^2c_{23} s_{12} s_{23} s_{13}\cos\gamma} .
 \label{eq:Cond11}
}
The second relation of (\ref{eq:Cond11}) is the important parameter in the neutrino oscillation and hence the condition   for the (11) element to reproduce the PDG's (11) element is intuitive and persuasive.  From the known values of $\theta_{ij}$ and the solutions of (\ref{eq:angles}), $\gamma$ should be bounded. Especially,    it cannot be $-\frac{\pi}{2}$. The numerical solutions for the angles in the KS form in the PMNS matrix will be presented  in Sec. \ref{sec:Mdiagonal}.

\section{Diagonalization of mass matrices and mixing angles in the KS form}\label{sec:Mdiagonal}

The charged current (CC) coupling in the lepton sector is  
\dis{
{\cal L}_{\rm CC}=- \frac{g}{\sqrt2}\sum_{l=e,\mu,\tau}\bar{l}_L\gamma^\alpha \nu_{lL}W_{\alpha}^-+{\rm h.c.}
}
where the weak eigenstate leptons $l$ are the defining ones in the CC interaction, and the weak eigenstate leptons $l, \nu_l$ are related to the mass eigenstate leptons $l^{\rm\, (mass)}_i,\nu_i^{\rm\, (mass)}$ as
\dis{
l_{L}=\sum_{j=1}^{3}V^{(e)}_{lj}l_{jL}^{\rm (mass)},~\nu_{lL}=\sum_{j=1}^{3}V^{(\nu)}_{lj}\nu_{jL}^{\rm\, (mass)}.\label{eq:LepUnitary}
}
Between the mass eigenstates, the CC interaction is given by
\dis{
{\cal L}_{\rm CC}=- \frac{g}{\sqrt2}\,\bar{l}_L^{\rm\, (mass)}\gamma^\alpha 
V^{(e)\dagger}V^{(\nu)}\nu_{L}^{\rm (mass)}W_{\alpha}^-+{\rm h.c.}
}
The PMNS matrix is given by\footnote{Compare with the CKM matrix $\Vckm=V^{(u)\dagger}V^{(d)}$ defined from the $W^+_\mu$ coupling, $- \frac{g}{\sqrt2}\,\bar{u}_L^{\rm\, (mass)}\gamma^\alpha 
V^{(u)\dagger}V^{(d)}d_{L}^{\rm (mass)}W_{\alpha}^++{\rm h.c.}$ } 
\dis{
V_{\rm PMNS}^\dagger =V^{(e)\dagger}V^{(\nu)}
}
where $V^{(e )}$ and $ V^{(\nu)}$ are diagonalizing unitary matricies of L-handed charged leptons and neutrino fields.

\begin{figure}[!t]
\includegraphics[width=0.55\textwidth]{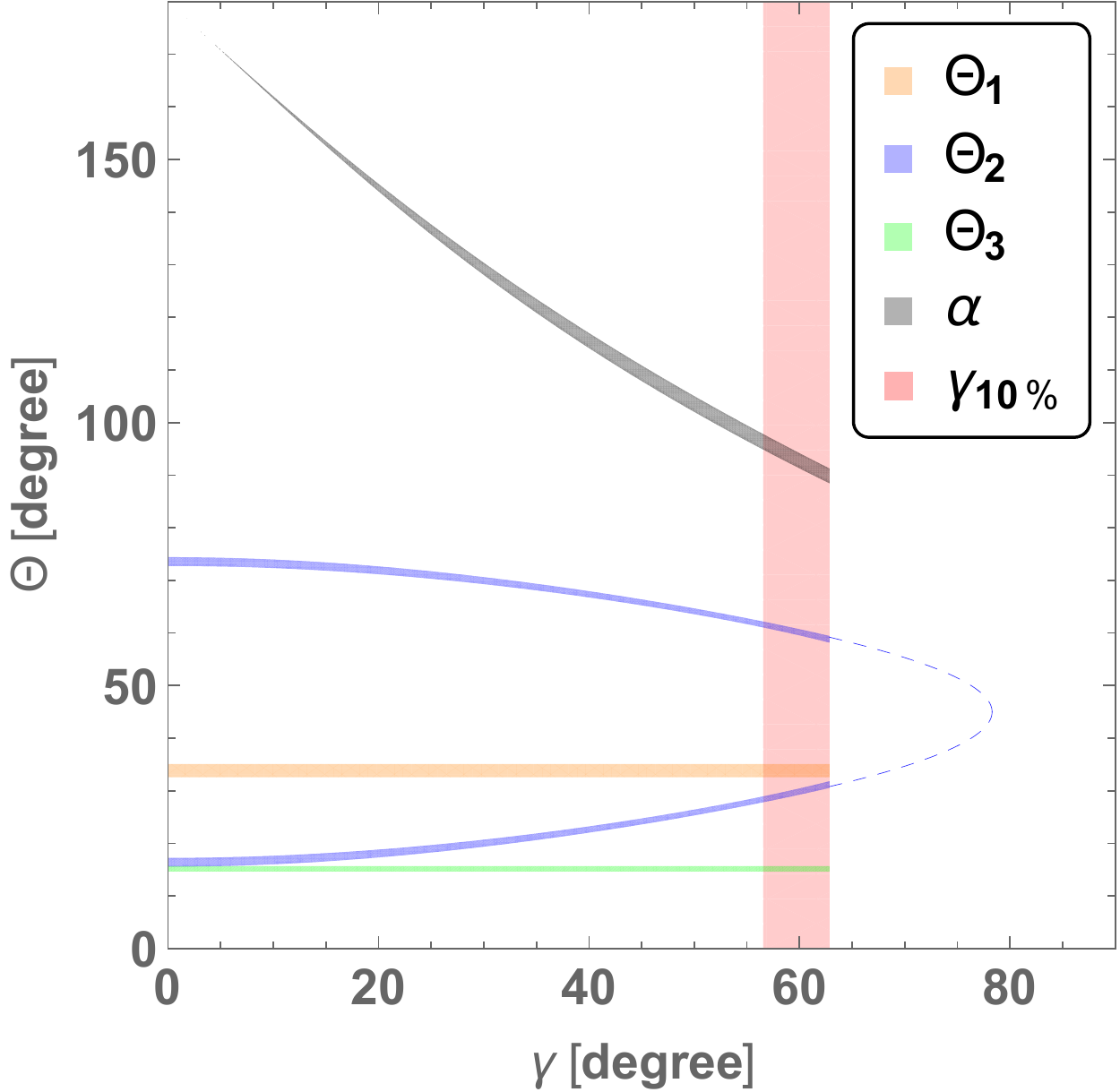} 
 \caption{The KS values of $\theta_{1,2,3}$ and $\alpha$ in the first quadrant within one sigma bounds of the PDG values as  functions in the allowed region of $\gamma$. In the second quadrant,  $\theta_{1,2,3}$ are given as  $\theta_{1,2,3}(-\gamma) =\theta_{1,2,3}(\gamma)$. In the third quadrant $\alpha$ is given as $\alpha(-\gamma)\simeq -\alpha(\gamma)$.  The 10\,\% allowed region from the maximum,   $\gamma=[+62.8^{\rm o}, +56.56^{\rm o}]$, is shown as the pink band. The solution for $\Theta_2$ is an ellipse whose tangent, as shown by the extended dashed curve, gives the limit determining the maximum $|\gamma|$'s in Eq. (\ref{eq:gammabd}).
} \label{fig:Thetas} 
\end{figure}

\begin{figure}[!t]
\includegraphics[width=0.55\textwidth]{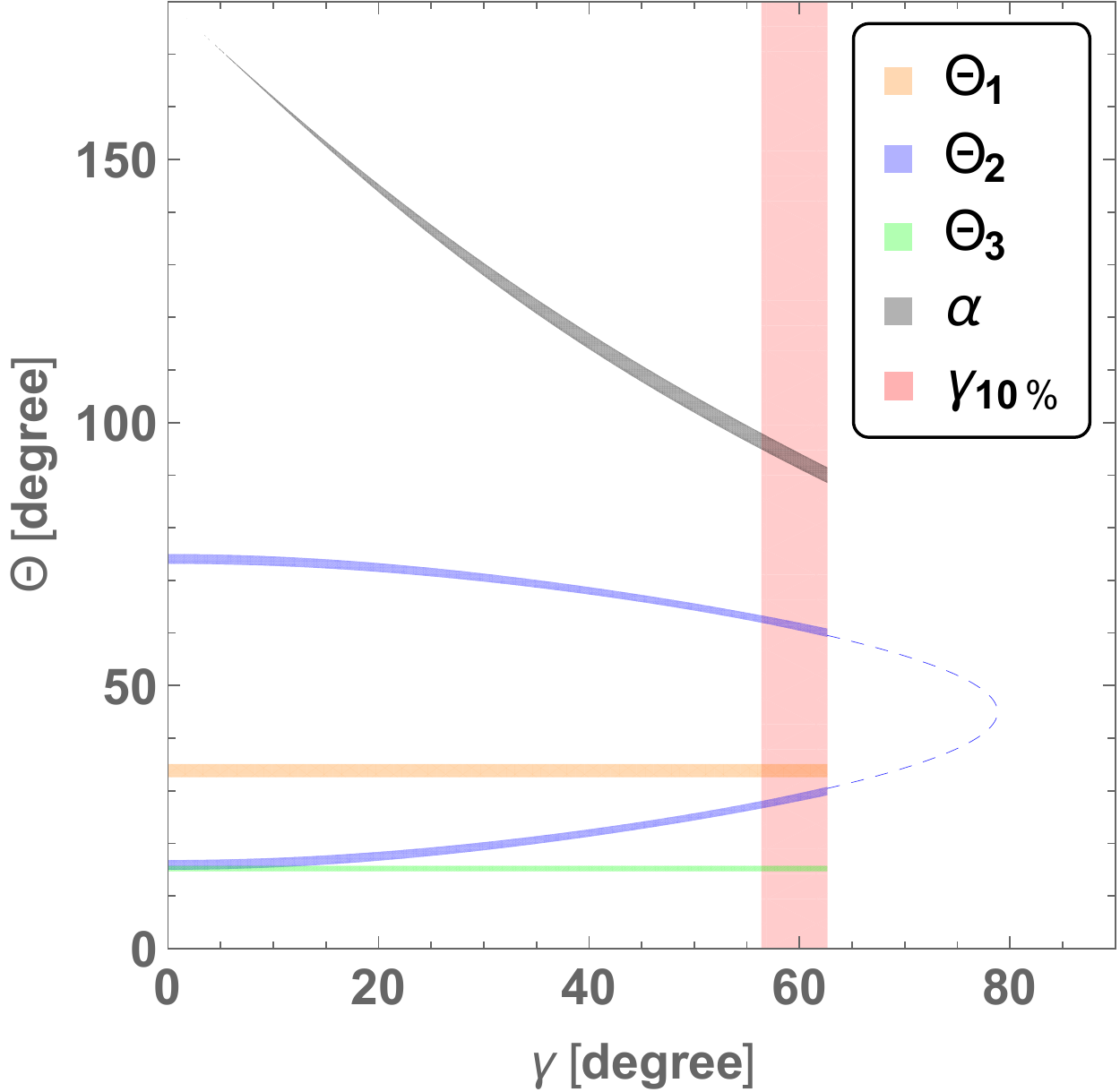} 
 \caption{Same as in Fig. \ref{fig:Thetas} except for the inverted hierarchy \cite{PDG18pmns}, $m_3<m_1<m_2$, where Eq. (\ref{eq:gammabd}) becomes $-78.80^{\rm o\, +0.94^{\rm o}}_{~\,-3.01^{\rm o}} <\gamma < +78.80^{\rm o\, +3.01^{\rm o}}_{~\,-0.94^{\rm o}} $ and the lower limit of $\gamma$ becomes $-62.64^{\rm o\, +1.46^{\rm o}}_{~\,-1.48^{\rm o}} <\gamma $.
} \label{fig:ThetasIn} 
\end{figure}

A standard way to parametrize the CC lepton interactions is
\dis{
{\rm CC ~lepton~matrix}=\Vpmns^\dagger\times \begin{pmatrix} 1,&0&0\\
0&e^{i \alpha_{21}/2}&0\\
0&0&e^{i \alpha_{31}/ 2}
\end{pmatrix}\label{eq:Majorana}
}
where the first factor called the PMNS matrix  is usually written as \cite{Valle80,PDG18pmns}
\dis{
\Vpmns^\dagger \simeq 
\begin{pmatrix}
C_{12} C_{13} , & S_{12} C_{13}  , & S_{13} e^{-i\delta} \\[0.5em]
-S_{12} C_{23} -C_{12}S_{23} S_{13} e^{i\delta}, & C_{12} C_{23} -S_{12}S_{23}S_{13} e^{i\delta}  , & S_{23}C_{13}   \\[0.5em]
S_{12} S_{23} -C_{12} C_{23}S_{13} e^{i\delta} , & -C_{12} S_{23}-S_{12} C_{23} S_{13} e^{i\delta}, & C_{23} C_{13}
\end{pmatrix}, \label{eq:CKMpdg18} 
}
where $C_{ij}=\cos\theta_{ij}, S_{ij}=\sin\theta_{ij}$,  $\theta_{ij}=[0,\frac{\pi}{2})$, and the angle $ \delta=[0,2\pi]$ is the Dirac CP violation phase, and $\alpha_{21}, \alpha_{31}$ are two Majorana CP violation phases. The second factor of (\ref{eq:Majorana}) contains the Majorana phases which may be determined by heavy neutrinos in the seesaw mechanism.
The best fit(BF) real angles  of the PMNS matrix   are \cite{PDG18pmns},
\dis{
&\Theta_{12}=0.5764~[C_{12}
=0.8385,\,S_{12}=0.5450 ],\\
&\Theta_{23}=0.7101~[C_{23}=0.7583 ,\, S_{23}=0.6519],\\
&\Theta_{13}=0.1472~[C_{13}=0.9892,\,S_{13}=0.1466 ].\label{eq:PMNSangles}
}
and we have the following bound from Fig. \ref{fig:Thetas},
\dis{
 -78.29^{\rm o\, +2.35^{\rm o}}_{~\,-0.90^{\rm o}} <\gamma < +78.29^{\rm o\, +0.90^{\rm o}}_{~\,-2.35^{\rm o}} 
 \label{eq:gammabd}
}  
from which we obtain
\dis{
\Vpmns^\dagger &\simeq 
\begin{pmatrix}
 0.8294 , & 0.5391, & 0.1466\, e^{-i\delta} \\[0.5em]
  -0.4132-0.08015\, e^{i\delta} ,& 0.6358-0.0521\, e^{i\delta}  , & 0.6449
  \\[0.5em]
 0.3553-0.0932  \, e^{i\delta}  , & -0.5466-0.0606\, e^{i\delta}  , & 0.7501
\end{pmatrix},\\[0.3em]
&J={\rm Im}\,V_{11}V_{22}V_{33}=  -3.28\times 10^{-2}\sin\delta\label{eq:PMNS18}
}  
where we used the central values for the allowed angles, $\theta_{12}=0.5758 (=32.99^{\rm o}), \theta_{13}=0.1471 (=8.428^{\rm o})$ and   $\theta_{23}=0.7101 (=40.69^{\rm o})$, for the normal hierarchy\footnote{Here  we cite, for simplicity of presentation, mainly the numbers for normal hierarchy of neutrino masses except in Fig. \ref{fig:ThetasIn}. } of neutrino masses  $m_1<m_2<m_3$.  

However, it is useful to have real numbers in  one row of the PMNS matrix as in \cite{KimSeo11},
\dis{
V_{\rm KS}^\dagger= \left(\begin{array}{ccc} C_1,&S_1C_3,&S_1S_3 \\ [0.2em]
 -C_2S_1,& C_1C_2C_3+S_2S_3e^{-i\delks} ,& C_1C_2S_3- S_2C_3e^{-i\delks}\\[0.2em]
-e^{i\delks} S_1S_2,&-C_2S_3 +C_1S_2C_3 e^{i\delks },& C_2C_3 +C_1S_2S_3 e^{i\delks}
\end{array}\right) ,\label{eq:KSWminus}
}
where Det$\,V_{\rm KS}=1 $. Then, the   phase appearing in the (31) element is the key, viz. $J= -{\rm Im\,}V^{\rm KS}_{31}V^{\rm KS}_{22}V^{\rm KS}_{13}=-C_1C_2C_3 S_1^2 S_2 S_3\sin\delks$ \cite{KimSeo12}.
 
 To make the  PMNS matrix with  one row  real from the numbers given in Eq. (\ref{eq:PMNSangles}),  we present  numerical solutions of Eq. (\ref{eq:Cond11}) in Fig. \ref{fig:Thetas}.\footnote{An approximate analytic solution near the dodeca symmetric point was given before \cite{KimSeoPMNS12}. } 
The BF real angles  from \cite{PDG18pmns} determine $\Theta_1$ and $\Theta_3$ accurately,
\dis{
&\Theta_{1}=0.5928  ~[ C_{1}
=0.8294,\,S_{1}=0.5587],\\ 
&\Theta_{3}=0.2656~[ C_{3}=0.9649,\, S_{3}=0.2625],\label{eq:PMNSangles}
}
but  $\Theta_2$ is can be 0.5377 or 1.0331. 
For $\alpha=-\frac{\pi}{2}$ (corresponding to $\gamma=-62.8^{\rm o} $) and $\Theta_2= 0.5377$, we have
\dis{
V_{\rm KS}^\dagger &= \left(\begin{array}{ccc} 0.82939,& 0.53909,&  0.14663 \\ [0.2em]
 -0.47985, & 0.68740+0.13441 e^{-i\delks} ,&
0.18697 -0.49417 e^{-i\delks}  \\[0.2em]
-0.28611 e^{i\delks} ,&-0.22543 +0.40986e^{i\delks}  ,& 0.82880 +0.11148 e^{-i\delks}
\end{array}\right),\\[0.3em]
& J= -{\rm Im\,}V^{\rm KS}_{31}V^{\rm KS}_{22}V^{\rm KS}_{13}=- 2.8838\times 10^{-2}\sin\delks\,.\label{eq:KSWnum}
}
Namely, to have $J$ given in Eq. (\ref{eq:KSWnum}) for $\delks=-\frac{\pi}{2}$ compared to $J$ of Eq. (\ref{eq:PMNS18}), we have the {\it minimum allowed value $\gamma=-62.8^{\rm o}$} which is inside  the region given in Eq. (\ref{eq:gammabd}).
In Fig.  \ref{fig:Thetas}, we mark the $+ 10\,\%$ band from this value,  $\gamma=[62.8^{\rm o}, 56.52^{\rm o}]$, as the pink band.  
In the third quadrant, the band becomes anti-symmetric to the curve in the first quadrant, $\gamma=[-62.8^{\rm o~+1.25^o}_{\rm~~-1.56^o}, -56.56]$. 
In Fig. \ref{fig:ThetasIn}, we present an  inverted hierarchy solution for $m_3<m_1<m_2$. 

We used $\dagger$ notation in (\ref{eq:KSWminus}) since the definition of the PMNS matrix is given by $W^-_\mu$ coupling and the CKM matrix is  given by $W^+_\mu$ coupling. To compare both with the $W^+_\mu$ coupling,  factoring out the  Majorana phases, let us consider the PMNS parametrization with  $\dagger$ of Eq. (\ref{eq:KSWminus}),
\dis{
V_{\rm KS}= \left(\begin{array}{ccc} C_1,&-C_2S_1,& -S_1S_2e^{-i\delks}  \\ [0.2em]
S_1C_3, & C_1C_2C_3+ S_2S_3 e^{i\delks},&
-C_2S_3 +C_1S_2C_3 e^{-i\delks}  \\[0.2em]
S_1S_3,&  C_1C_2S_3-S_2C_3e^{i\delks}  ,& C_2C_3 +C_1S_2S_3 e^{-i\delks}
\end{array}\right),\label{eq:KSWplus}
}
To build a model, leading to  (\ref{eq:KSWplus}), one must find out the mass matrices $M^{(\nu)}$ and $M^{(l)}$ with appropriate insertions of $e^{\pm i\delks}$.
\begin{figure}[!t] 
\includegraphics[width=0.43\textwidth]{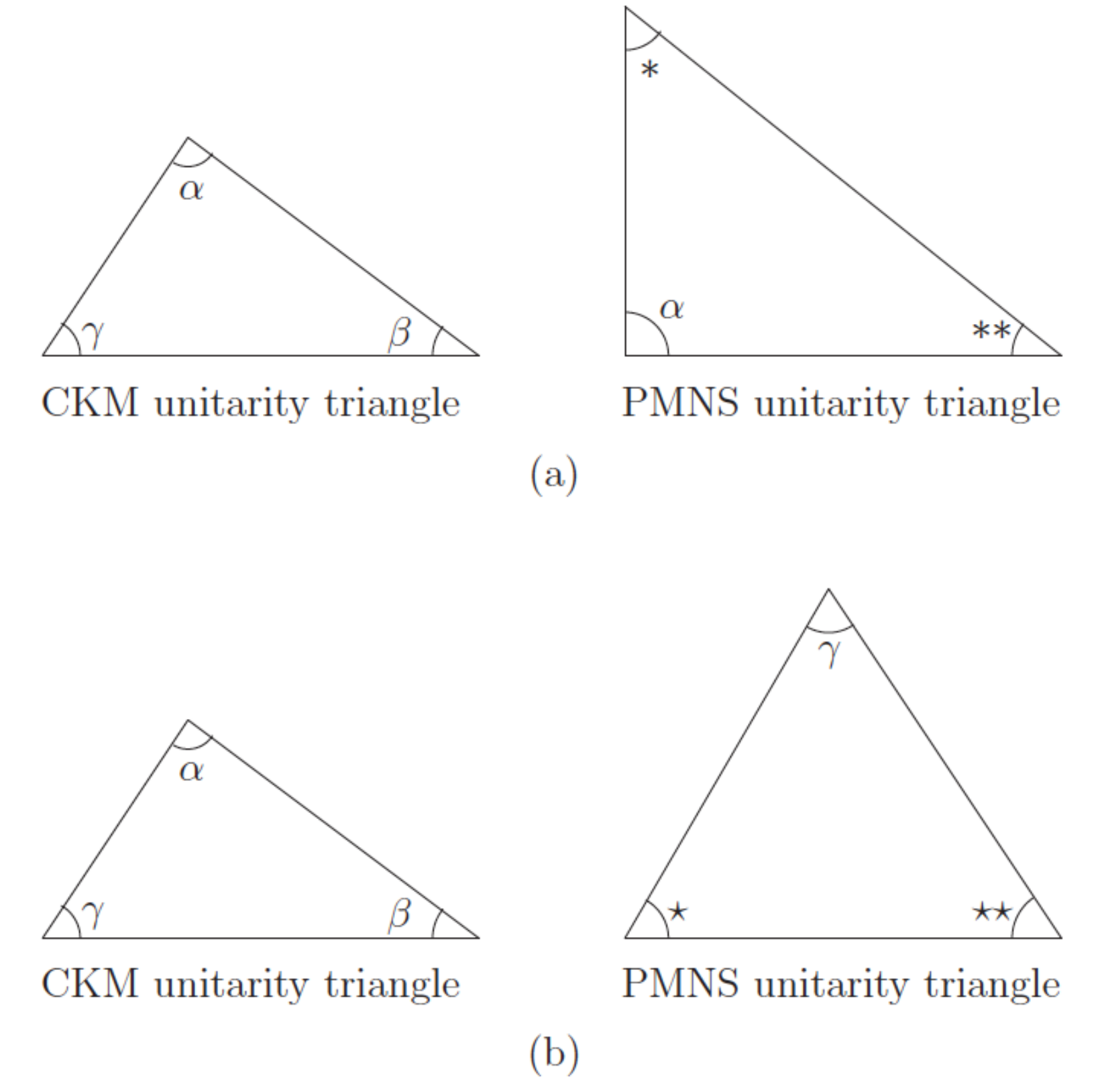} 
   \caption{The CKM and PMNS unitary triangles with one common angle \cite{NamCKMPMNS}:
  (a) $\alpha$ in the KS  form, and (b) $\gamma$ in the SV form.} \label{fig:KSsv} 
\end{figure}

As suggested in \cite{NamCKMPMNS}, if we use the KS parametrization for the CKM matrix given in \cite{KimSeo11} and again the KS parametrization for the PMNS matrix \cite{KimSeoPMNS12} given in Eq. (\ref{eq:KSWplus}) and the same CP phase $\alpha$ appears in the CKM and PMNS phases, we expect the unitary triangles take the forms given in Fig. \ref{fig:KSsv} \,(a).  If we use the Maiani-Chau-Keung (MCK) parametrization for the CKM matrix   and the  Schechter-Valle (SV) parametrization for the PMNS matrix given  in \cite{Valle80,PDG18pmns} and the same CP phase $\gamma$ appears in the CKM and PMNS phases, we expect the unitary triangles take the forms given  in Fig. \ref{fig:KSsv}\,(b). The CKM unitary triangle is known rather accurately but the PMNS unitary triangle   is not known accurately, chiefly because the error bars allowed for $\gamma$ is large: e.g. for the normal hierarchy  $\delta_{\rm CP}= -1.728^{+0.851}_{-0.855}$  \cite{T2K18}. These unitary triangles are defined by CC interactions, and determined chiefly by the decay processes in the quark sector and by neutrino oscillations in the lepton sector.
  
\section{Suggestion from the flpped SU(5) model}\label{sec:FSU5}

\begin{figure}[!h]
\includegraphics[width=0.35\textwidth]{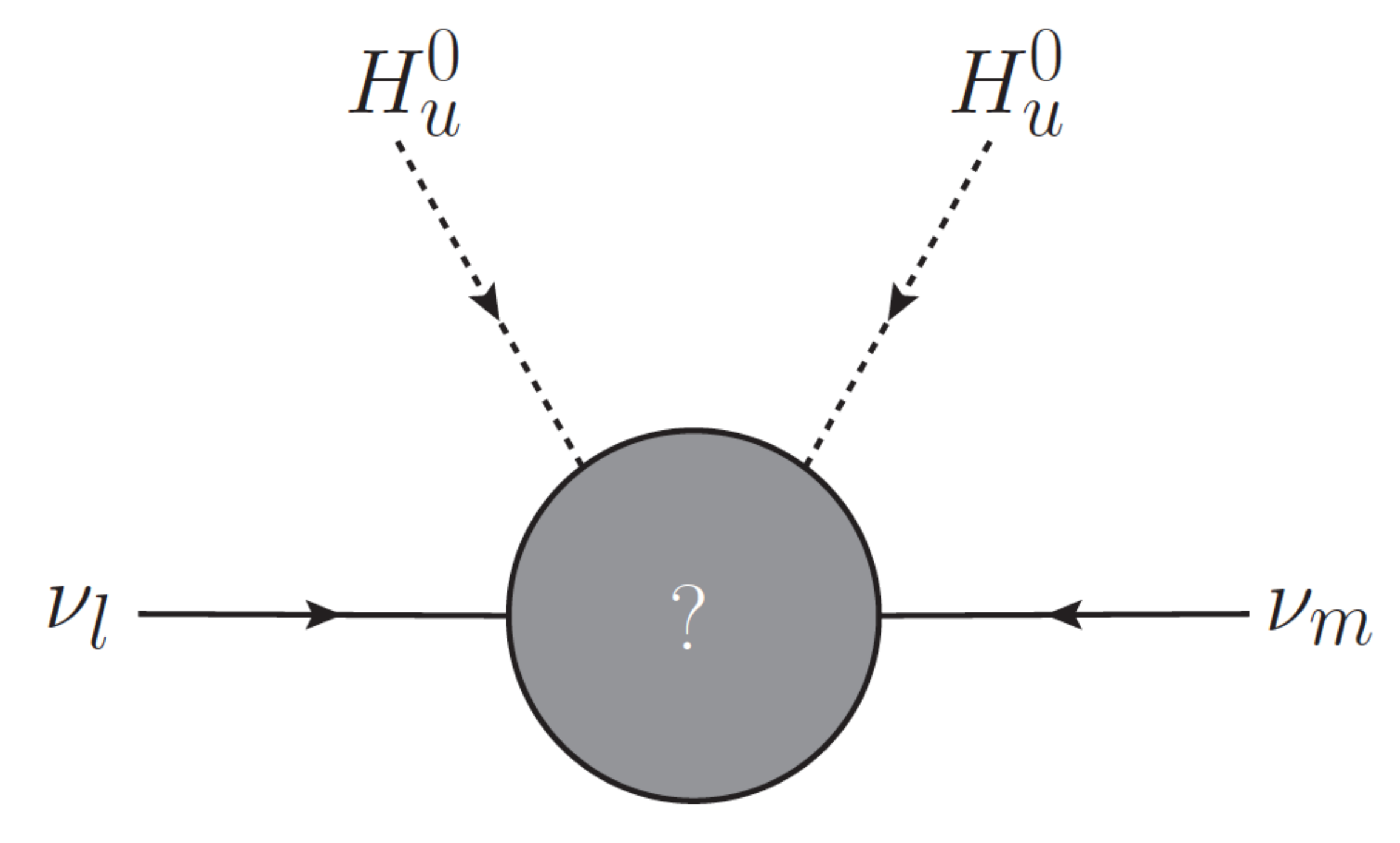} 
   \caption{A neutrino interaction with the SM fields only.} \label{fig:NuMass} 
\end{figure}

If we consider only the SM particles, neutrino masses arise from the diagram shown in Fig.  \ref{fig:NuMass}. Any further attachments to this diagram are  SM singlet scalars. If we consider the quantum numbers under SU(2)$_W\times$U(1)$_Y$, two neutrinos have ${\one}_{-1}\oplus {\three}^\uparrow _{-1}$ where $\uparrow$ means that the 3rd component of the weak isospin is $+1$. Possible scalar attachments must carry quantum number 
${\one}_{+1}$ or $ {\three}^\downarrow _{+1}$, and ${\one}_{+1}$ is ruled out because  $\langle{\one}_{+1}\rangle$ breaks U(1)$_{\rm em}$.  $ {\three}^\downarrow _{+1}$ allows the scalar attachments, shown as $H_u\oplus H_u$ in  Fig.  \ref{fig:NuMass}.  Depending on details of high energy fields, implied by the question mark in the gray, two types of neutrino masses are named, Type I seasaw \cite{TypeI} and Type II seasaw \cite{TypeII}. Type III seasaw \cite{TypeIII} requires more light particles at the electroweak scale. From the \flip~spectra shown in Ref. \cite{KimKyae17}, we note that there is no SU(2) triplet representation;  hence only Type I seasaw is allowed from our string compactification.

Considering the SM singlet attachements to  Fig.  \ref{fig:NuMass}, let us consider
the neutrino mass operator allowed in the $\Z_{12-I}$ compactification. Firstly, the diagonal masses are
\dis{
M^\nu_{33} &\propto\frac{1}{\tilde{M}_3^3}\int d^2\vartheta   \, \five_{+3}(U_3,0;+1)\five_{+3}(U_3,0;+1) \five_{-2}(T_6,\frac13;-2)  \five_{-2}(T_6,\frac13;-2)\tenb_{-1}(T_3,\frac13;+4)\tenb_{-1}(T_3,0;+4) \\
M^\nu_{22} &\propto \frac{1}{\tilde{M}_2^4} \int d^2\vartheta d^2\bar{\vartheta}  \,\five_{+3}(T_4^0,\frac14;-1)\five_{+3}(T_4^0,\frac14;-1)    \five_{-2}(T_6,\frac13;-2)  \five_{-2}(T_6,\frac13;-2) \\
&\qquad\qquad \cdot\tenb_{-1}(T_3,\frac13;+4)\tenb_{-1}(T_3,0;+4) \one_{0}( \sigma_5,T_6,\frac12;+4)\\
 \label{eq:nuMass}
}
where the last number after ; is the $Q_R$ charge, and $\tilde{M}_3$ and  $\tilde{M}_2$ are determined by ? in Fig. \ref{fig:NuMass}. We need $Q_R=2$ modulo 4 for $d^2\vartheta$ integration and  $Q_R=0$  modulo 4 for $d^2\vartheta d^2\bar{\vartheta}$ integration. $M^\nu_{11}, M^\nu_{12}, M^\nu_{21}$ have the same structure as $M^\nu_{22}$. The quantum numbers are listed in Tables I and II. Note that the selection rule is making the phase an  integer multiple, which is satisfied above, viz.   $\frac13+\frac13+\frac13=1 $ and $\frac14+\frac14+\frac13+\frac13+\frac13+\frac12=2$.
Then, the above  masses are estimated as 
\dis{
M^\nu_{33}\sim  \frac{ v_{EW}^2 M_{10}^2 }{\tilde{M}_3^3}, ~M^\nu_{22}\sim   \frac{v_{EW}^2 M_{10}^2 | \langle\sigma_5  \rangle |}{\tilde{M}_2^4}  .\label{eq:nuOrder}
}
Then, neutrino mixing masses are generally of order $v_{EW}^2/\tilde{M}$ since the SM singlet VEVs can be at the GUT scale without breaking $\Z_{4R}$.

For the off-diagonal masses  between $U_3$ and $T_4^0$ neutrinos, we need $d^2\vartheta d^2\bar{\vartheta}$ integration,
\dis{
M_{32}^\nu, M^\nu_{31}\propto &\frac{1}{ \tilde{M}^6 }\int d^2\vartheta d^2\bar{\vartheta} \,  \five_{+3}(U_3,0;+1)\five_{+3}(T_4^0,\frac14;-1)   \five_{-2}(T_6,\frac13;-2)  \five_{-2}(T_6,\frac13;-2)\\
& \qquad \qquad \cdot\tenb_{-1}(T_3,0;+4) \tenb_{-1}(T_3,0;+4)   \cdot  \one_0(\sigma_1,T_4^0,\frac14;-4) ^*\one_0(\sigma_{14},T_3,\frac23;-4) ^*.
}
Then, the above  mass mixing is estimated as 
\dis{
M^\nu_{13,23}\sim    \frac{v_{EW}^2  M_{10}^2 | \langle\sigma_1  \sigma_{14} \rangle |}{\tilde{M}_{\times}^5},\label{eq:nu23mixing}
}
where $\tilde{M}$ is some mass scale determined by the above equations.
Note that $\Sigma_2^*,\Sigma_1$ and $\sigma_1$  can have the GUT scale VEVs because all of them carry $Q_R=0$ modulo 4, and we obtain a similar order of mass for all of $M^\nu_{11,12,22,33,13,23}$.  
 
 Comparing   $M^\nu_{11,22,31,32}$ and $M^\nu_{33}$, 
\dis{
\frac{M^\nu_{11},M^\nu_{22}}{M^\nu_{33}}\approx \left|\frac{\sigma_{5} }{\tilde{M}} \right|,
\frac{M^\nu_{31},M^\nu_{32}}{M^\nu_{33}}\approx \left|\frac{\langle\sigma_{1}\sigma_{14}\rangle }{\tilde{M}^2} \right|,
}
we note that the neutrino mass hierarchy favors the normal hierarchy (in the sense that $\nu_\tau$ is the heaviest) if the VEVs of $\sigma$ singlets are comparably small, $|\sigma_{1}|,|\sigma_{5}| <\tilde{M}$.
 
 Since we obtained all entries in the neutrino mass matrix, here we investigate how the CP phase can be inserted in the mass matrix of the $\Qem=-1$ leptons   and in the neutrino mass matrix.
 
\subsection{Neutrino mass matrix  inspired by flipped SU(5)}\label{subsec:ModelNu}

In Ref. \cite{Kim18Rp} based on the flipped SU(5) model of 
\cite{Huh09}, a possible identification $\Z_{4R}$ has been achieved, forbidding dimension-5 B violating operators but allowing the electroweak scale $\mu$ term and dimension-5 L violating Weinberg operator. The $\Z_{4R}$ quantum numbers,   $Q_{4R}$, of   the SM fields and  neutral singlets ($\sigma$'s), are presented  in Ref. \cite{Kim18Rp}.
In the flipped SU(5), the  neutrino masses arise in the form
\dis{
-{\cal L}_{\nu}^{IJ}=f_{IJ}^{(\nu)}(\{\sigma\})  \five_{+3}^{I,i}\five_{+3}^{J,j}\five_{-2}^{k}(H_u)\five_{-2}^{l}(H_u) [\tenb_{-1}(H_{\rm GUT}) \tenb_{-1}(H_{\rm GUT})]_{ijkl}+{\rm h.c.}, \label{eq:dYukawa}
}
where the couplings $f_{IJ}^{(\nu)} $ are complex parameters, $I$ and  $J$ are flavor indices,  $i,j,k,l,m$ are SU(5) indices, and the subscript is the U(1)$_X$ quantum number of \flip. $\five_{-2}$ is usually denoted as $H_{uL}$, and $ \tenb_{-1}$, together with $\ten_{+1}$, is the ten-plet needed for breaking the rank 5 gauge group SU(5)$\times$U(1) at a GUT scale down to the rank 4 SM gauge group. These quantum numbers in \flip\,are  given in  Ref. \cite{Kim18Rp}.   

Consider $ \five_{+3}^{I,i}\five_{+3}^{J,j}$ in Eq. (\ref{eq:dYukawa}) which   is symmetric under $I$ and $J$.  Thus, the neutrino mass matrix is symmetric. The Majorana phase factored in Eq. (\ref{eq:Majorana}) is from the heavy neutrinos, which does not affect our satudy of CC interactions shown in Eq. (\ref{eq:CKMpdg18}).  As in the quark case, we assume that the neutrino mass matrix determining the PMNS matrix is real.  Thus, $V^{(\nu)}$ can be considered to be an orthogonal matrix $O^{(\nu)}$.

\subsection{Mass matrix of charged leptons  inspired by flipped SU(5)}\label{sec:ModelEl}
We can always take $U^{(\nu)}$ as a real matrix $O^{(\nu)}$. Thus, the PMNS matrix given in (\ref{eq:KSWminus}) can be represented as
\dis{
V_{\rm KS}^{(l)\dagger}= V^{(e)}O^{(\nu)T}=\begin{pmatrix}  q_{11} r_{11} + q_{12} r_{12} + q_{13} r_{13}, &   q_{11} r_{21} + q_{12} r_{22} + q_{13} r_{23},&  q_{11} r_{31} + q_{12} r_{32} + q_{13} r_{33},\\
 q_{21} r_{11} + q_{22} r_{12} + q_{23} r_{13}, &   q_{21} r_{21} + q_{22} r_{22} + q_{23} r_{23},&  q_{21} r_{31} + q_{22} r_{32} + q_{23} r_{33},\\
 q_{31} r_{11} + q_{32} r_{12} + q_{33} r_{13}, &   q_{31} r_{21} + q_{32} r_{22} + q_{33} r_{23},&  q_{31} r_{31} + q_{32} r_{32} + q_{33} r_{33},\\
\end{pmatrix}  \label{eq:KSelem1}
}
where the elements  $V_{ij}^{(l)}=q_{ij} $ and $O_{ij}^{(\nu)}=r_{ij} $ are complex  and real numbers numbers, respectively. Comparing with Eq. (\ref{eq:KSWminus}), $q_{11},q_{12}$ and $q_{13}$ are required to be real.

 The unitary matrices relating the weak eigenstates $l$ amd mass eigenstates $i $ of the charged leptons are named as $V$ for L-handed fields and $U$ for R-handed fields,
\dis{
l_{L}=\sum_{j=1}^{3}V^{(l)}_{lj}l_{jL}^{\rm (mass)},~l_R=\sum_{j=1}^{3}U^{(l)}_{lj}l_{jR}^{\rm\, (mass)}.\label{eq:LRrel}
}
The mass matrix $\bar{l}^{\,\rm mass}_L\, {\rm diag. }\,(\tilde{m}_e, \tilde{m}_\mu,1)l_R^{\rm mass}$ (where $\tilde{m}_l=m_l/m_\tau$) in the mass eigenstate basis becomes
\dis{
\bar{l}_L(V_e^{(l)}M^{\rm mass}U_e^{(l)\dagger} )l_R
}
in the weak eigenstate basis.  Since R-handed leptons are not participating in the CC interactions,  the lepton R-handed unitary matrix $U^{(l)}$ can be taken as the identity matrix. Thus, the mass matrix in the weak basis becomes
\dis{
M^{(l)} &=V^{(l)}\begin{pmatrix} \tilde{m}_e,&0,&0\\ 0,&\tilde{m}_\mu,&0\\ 0,&0,&1
\end{pmatrix}U^{(l)\dagger}= \begin{pmatrix} q_{11}\tilde{m}_e ,& q_{12} \tilde{m}_\mu, & q_{13}  \\
q_{21}\tilde{m}_e ,& q_{22} \tilde{m}_\mu, & q_{23}  \\
q_{31}\tilde{m}_e ,& q_{32} \tilde{m}_\mu, & q_{33}  \\
 \end{pmatrix} = \begin{pmatrix} {\rm real} ,&  {\rm real} , &  {\rm real}  \\
 {\rm complex}  ,&  {\rm complex}, & {\rm complex}  \\
 {\rm complex} ,&  {\rm complex}, &  {\rm complex}\\
 \end{pmatrix}  \label{eq:Mweak1}
}
where   $V_{ij}^{(l)}=q_{ij} $  and we obtained  $q_{11},q_{12}$ and $q_{13}$ are real numbers.
 
We show that the quantum numbers of the model presented in \cite{Kim18Rp} allows an effective mass matrix form  Eq. (\ref{eq:Mweak1}) for the charged leptons.
\dis{
-{\cal L}_{l}^{IJ}=f_{IJ}^{(l)}(\{\sigma\})  \five_{+3}^{I,i}\one_{-5}^{J}\fiveb_{+2 ,i}(H_d)  [\ten_{+1}(H_{\rm GUT}) \tenb_{-1}(H_{\rm GUT})] +{\rm h.c.}, \label{eq:dYukawa}
}
which arises from, for example for the (22), (33) and (32) elements, viz. Tables II and III, 
\dis{
 & \frac{1}{ {M}^4}\int d^2\vartheta d^2\bar\vartheta\, \bar{\eta}_2(T_4^0,\frac14) \mu^c(T_4^0,\frac14) H_d(T_6,\frac13)
\Sigma_2(T_3,0) \Sigma_1^*(T_3,\frac23)\sigma_5^*(T_6,-\frac12)\\
 & \frac{1}{ {M}^4}\int d^2\vartheta d^2\bar\vartheta\, \bar{\eta}_3(U_3,0) \tau^c(U_3,0) H_d(T_6,\frac13)
\Sigma_2(T_3,0) \Sigma_1^*(T_3,\frac23)\sigma_2^*(T_4^0,0)\\
& \frac{1}{ {M}^4}\int d^2\vartheta  \, \bar{\eta}_3(U_3,0) \mu^c(T_4^0,\frac14) H_d(T_6,\frac13)
\Sigma_2(T_3,0) \Sigma_1^*(T_3,\frac23)\sigma_1(T_4^0, \frac14)\sigma_5(T_6,\frac12)  \label{eq:effMe1}
}
which is allowed with $Q_R=0$ (needed for D-terms) and $Q_R=2$ (needed for F-terms) modulo 4, respectively. The BSM fields in Eq. (\ref{eq:effMe1}) carry $Q_R=4$ and $\Z_{4R}$ is not broken by the mass terms of the charged leptons.
 
\begin{table}[t!]
\begin{center}
\begin{tabular}{@{}lc|cc|c|cccccc|c|ccc@{}} \toprule
 &State($P+kV_0$)&$~\Theta_i~$ &${\bf R}_X$(Sect.)&$Q_R$ &$Q_1$&$Q_2$ &$Q_3$ &$Q_4$ &$Q_5$ &$Q_6$ & $Q_{\rm anom}$& $Q_{18}$& $Q_{20}$& $Q_{22}$ \\[0.1em] \colrule
 $\xi_3$  & $(\underline{+++--};--+)(0^8)'$&$0$ &$\tenb_{-1}(U_3)$&$+1$ &$-6$ & $-6$ & $+6$ & $0$ & $0$ & $0$ & $-13$ & $+1$ & $-1$& $+1$ \\
$\bar{\eta}_3$  & $(\underline{+----};+--)(0^8)'$&$0$ &$\five_{+3}(U_3)$&$+1$  & $+6$ & $-6$ & $-6$ & $0$ & $0$ & $0$ & $-1$ & $+1$& $-1$& $+1$ \\
$\tau^c$  & $({+++++};-+-)(0^8)'$& $0$ &$\one_{-5}(U_3)$ &$+1$& $-6$ & $+6$ & $-6$ & $0$ & $0$ & $0$ & $+5$ & $+1$& $-1$ & $+1$  \\
$\xi_2$  & $(\underline{+++--};-\frac{1}{6},-\frac{1}{6},-\frac{1}{6})(0^8)'$& $\frac{+1}{4}$ &$\tenb_{-1}(T_4^0)$& $-1$ &$-2$ & $-2$ & $-2$ & $0$ & $0$ & $0$ & $-3$ & $-1$& $-1$ & $-1$\\
$\bar{\eta}_2$  & $(\underline{+----};-\frac{1}{6},-\frac{1}{6},-\frac{1}{6})(0^8)'$&$\frac{+1}{4}$ &$\five_{+3}(T_4^0)$& $-1$   & $-2$ & $-2$ & $-2$ &$0$ & $0$ & $0$ & $-3$ & $-1$& $-1$ & $-1$\\
$\mu^c$  & $({+++++};-\frac{1}{6},-\frac{1}{6},-\frac{1}{6})(0^8)'$&$\frac{+1}{4}$ & $\one_{-5}(T_4^0)$& $-1$  & $-2$ & $-2$ & $-2$ &$0$ & $0$ & $0$ & $-3$ & $-1$& $-1$ & $-1$\\
$\xi_1$  & $(\underline{+++--};-\frac{1}{6},-\frac{1}{6},-\frac{1}{6})(0^8)'$&$\frac{+1}{4}$ &$\tenb_{-1}(T_4^0)$& $-1$  &$-2$ & $-2$ & $-2$ & $0$ & $0$ & $0$ & $-3$ &  $-1$& $-1$ & $-1$\\
$\bar{\eta}_1$  & $(\underline{+----};-\frac{1}{6},-\frac{1}{6},-\frac{1}{6})(0^8)'$&$\frac{+1}{4}$ &$\five_{+3}(T_4^0)$& $-1$   &$-2$ & $-2$ & $-2$ & $0$ & $0$ & $0$ & $-3$ & $-1$& $-1$ & $-1$\\
$e^c$  & $({+++++};-\frac{1}{6},-\frac{1}{6},-\frac{1}{6})(0^8)'$&$\frac{+1}{4}$ & $\one_{-5}(T_4^0)$& $-1$  &$-2$ & $-2$ & $-2$ &$0$ & $0$ & $0$ & $-3$ & $-1$& $-1$ & $-1$\\[0.2em]
\hline
 $H_{uL}$  & $(\underline{+1\,0\,0\,0\,0};\,0\,0\,0)(0^5;\frac{-1}{2}\,\frac{+1}{2}\,0)'$&$\frac{+1}{3}$ &$2\cdot \five_{-2}(T_6)$& $-2$  & $0$ & $0$ & $0$ & $-12$ & $0$ & $0$ & $0$ & $-1$& $-1$& $-1$   \\
 $H_{dL}$  & $(\underline{-1\,0\,0\,0\,0};\,0\,0\,0)(0^5;\frac{+1}{2}\,\frac{-1}{2}\,0 )'$&$\frac{+1}{3}$ &$ 2\cdot \fiveb_{+2}(T_6)$&$+2$ &$0$ & $0$ & $0$ & $+12$ & $0$ & $0$ & $0$ & $-1$ & $-1$& $-1$   \\ [0.2em]
 \botrule
 \end{tabular} \label{tab:SMfields} 
\end{center}
\caption{U(1) charges of matter fields in the SM.  $\xi_i$ and $\bar{\eta}_i$ contain  the left-handed quark and lepton doublets, respectively, in the i-th family.}
\end{table}

\begin{table}[t!]
\begin{center}
\begin{tabular}{@{}lc|ccc|c|cccccc|c|ccc@{}} \toprule
 &State($P+kV_0$) &$\Theta_i$ & $(N^L)_j$&${\cal P}\cdot {\bf R}_X$(Sect.)&$Q_R$ &$Q_1$&$Q_2$ &$Q_3$ &$Q_4$ &$Q_5$ &$Q_6$ & $Q_{\rm anom}$& $Q_{18}$& $Q_{20}$& $Q_{22}$  \\[0.1em] \colrule
 $\Sigma^*_1$  & $(\underline{+++--};0^3)(0^5;\frac{-1}{4}\,\frac{-1}{4}\,\frac{+2}{4})'$ &$0$ &$2(1_1) $ & $2\,\tenb_{-1}(T_3)_L$ &$+4$& $0$ & $0$ & $0$ & $0$ & $+9$ & $+3$ & $\frac{-33}{7}$ & $-1 $& $+1 $ & $-1 $  \\
$\Sigma^*_1$  & $(\underline{+++--};0^3)(0^5;\frac{-1}{4}\,\frac{-1}{4}\,\frac{+2}{4})'$ &$\frac{+2}{3}$ &$1(1_{3})$ & $1\,\tenb_{-1}(T_3)_L$ &$+4$& $0$ & $0$ & $0$ & $0$ & $+9$ & $+3$ & $\frac{-33}{7}$ & $-1 $& $+1 $ & $-1 $  \\
${\Sigma}_2$  & $(\underline{++---};0^3)(0^5;\frac{+1}{4}\,\frac{+1}{4}\,\frac{-2}{4})'$& $0$ &$2(1_{\bar{1}})$&$2\, \ten_{+1}(T_3)_L$ &$-4$&$0$ & $0$ & $0$ & $0$ & $-9$ & $-3$ & $\frac{+33}{7}$ & $-1  $& $-1 $ & $-1 $ \\
 ${\Sigma}_2$  & $(\underline{++---};0^3)(0^5;\frac{+1}{4}\,\frac{+1}{4}\,\frac{-2}{4})'$& $\frac{+1}{3}$ &$1(1_{\bar{3}})$&$1\, \ten_{+1}(T_3)_L$ &$-4$&$0$ & $0$ & $0$ & $0$ & $-9$ & $-3$ & $\frac{+33}{7}$ & $-1  $& $-1 $ & $-1 $ 
 \\[0.2em]
\hline
$\sigma_1$  & $(0^5;\frac{-2}{3}\, \frac{-2}{3}\, \frac{-2}{3})(0^8)'$&$\frac{+1}{4}$ & $0$&$2\cdot\one_{0}(T_4^0)$  &$-4$& $-8$ & $-8$ & $-8$ & $0$ & $0$ & $0$ & $-12$ & $-1  $& $-1 $ & $-1 $  \\
$\sigma_2$  & $(0^5;\frac{-2}{3}\, \frac{+1}{3}\, \frac{+1}{3})(0^8)'$&$0$ &$3(1_{\bar{1}})$& $3\cdot \one_{0}(T_4^0)$ &$0$& $-8$ & $+4$ & $+4$ & $0$ & $0$ & $0$ & $-2$ &  $-1  $& $-1 $ & $-1 $  \\
$\sigma_3$  & $(0^5;\frac{1}{3}\, \frac{-2}{3}\, \frac{1}{3})(0^8)'$&$0$ &$3(1_{\bar{1}})$&$3\cdot\one_{0}(T_4^0)$ &$0$&$+4 $ & $-8$ & $+4$ & $0$ & $0$ & $0$ & $-8$ &    $-1  $& $-1 $ & $-1 $  \\
$\sigma_4$  & $(0^5;\frac{1}{3}\, \frac{1}{3}\, \frac{-2}{3})(0^8)'$&$0$ &$3(1_{\bar{1}})$&$3\cdot\one_{0}(T_4^0)$  &$0$& $+4$ & $+4$ & $-8$ & $0$ & $0$ & $0$ & $+10$ &   $-1  $& $-1 $ & $-1 $  \\
$\sigma_5$  & $(0^5;0\,1\,0)(0^5;\frac{1}{2}\,\frac{-1}{2}\,0)'$&$\frac{+1}{2}$ &$0$& $2\cdot\one_{0}(T_6) $ &$+4$& $0$ & $+12$ & $0$ & $+12$ & $0$ & $0$ & $+14$ &   $-1  $& $-1 $ & $-1 $  \\
$\sigma_6$  & $(0^5;0\,0\,1)(0^5;\frac{-1}{2}\,\frac{1}{2}\,0)'$&$\frac{+1}{2}$ &$0$&$2\cdot\one_{0}(T_6) $ &$0$&$0 $ & $0$& $+12$ & $-12$ & $0$ & $0$ & $-4$ &  $-1  $& $-1 $ & $-1 $  \\
$\sigma_7$  & $(0^5;0\,-1\,0)(0^5;\frac{-1}{2}\,\frac{1}{2}\,0)'$&$\frac{+1}{2}$ &$0$&$2\cdot\one_{0}(T_6)_R$  &$+4$& $0$ & $+12$ & $0$ & $+12$ & $0$ & $0$ & $+14$ &   $-1  $& $+1 $ & $-1 $  \\
$\sigma_8$  & $(0^5;0\,0\,-1)(0^5;\frac{1}{2}\,\frac{-1}{2}\,0)'$&$\frac{+1}{2}$ &$0$& $2\cdot \one_{0}(T_6)_R$ & $-2$ & $0$&$0$ & $+12$ & $-12$ & $0$ & $0$ & $-4$ &   $-1  $& $+1 $ & $-1 $  \\
$\sigma_{11}$ & $(0^5;\frac{-1}{2}\,\frac{-1}{2}\,\frac{-1}{2})(0^5;\frac{3}{4}\,\frac{-1}{4}\,\frac{-1}{2})'$& $\frac{+2}{3}$ & $2(1_1+1_3,1_{\bar{1}}+ 1_{\bar{3}} )$& $ 2\cdot\one_{0}(T_3)$ &$-6$& $-6$ & $-6$ & $-6$ & $+12$ & $-9$ & $-3$ & $\frac{-30}{7}$ &  $+1  $& $+1 $ & $-1 $  \\
$\sigma_{11}'$ & $(0^5;\frac{-1}{2}\,\frac{-1}{2}\,\frac{-1}{2})(0^5;\frac{3}{4}\,\frac{-1}{4}\,\frac{-1}{2})'$&$0$ &$ 4(1_1+1_3,1_{\bar{1}}+ 1_{\bar{3}} )$& $  4\cdot \one_{0}(T_3)$ &$-6$& $-6$ & $-6$ & $-6$ & $+12$ & $-9$ & $-3$ & $\frac{-30}{7}$ &  $-1  $& $+1 $ & $+1 $  \\
   $\sigma_{12}$ & $(0^5;\frac{-1}{2}\,\frac{1}{2}\,\frac{1}{2}\,)(0^5;\frac{3}{4}\,\frac{-1}{4}\,\frac{-1}{2})'$&$\frac{+1}{3}$ & $2(1_1+1_3,1_{\bar{1}}+ 1_{\bar{3}} )$ &$ 2\cdot \one_{0}(T_3)$  &$-2$& $-6$ & $+6$ & $+6$ & $+12$ & $-9$ & $-3$ & $\frac{+40}{7}$ &  $+1  $& $+1 $ & $-1 $  \\
   $\sigma_{12}'$ & $(0^5;\frac{-1}{2}\,\frac{1}{2}\,\frac{1}{2}\,)(0^5;\frac{3}{4}\,\frac{-1}{4}\,\frac{-1}{2})'$&$\frac{+2}{3}$ & $2(1_1+1_3,1_{\bar{1}}+ 1_{\bar{3}} )$ &$ 2\cdot  \one_{0}(T_3)$  &$-2$& $-6$ & $+6$ & $+6$ & $+12$ & $-9$ & $-3$ & $\frac{+40}{7}$ &  $-1  $& $+1 $ & $+1 $  \\
  $\sigma_{13}$ & $(0^5;\frac{ 1}{2}\,\frac{1}{2}\,\frac{-1}{2}\,)(0^5;\frac{-1}{4}\,\frac{3}{4}\,\frac{-1}{2})'$&$\frac{+1}{3}$ &$2(1_1+1_3,1_{\bar{1}}+ 1_{\bar{3}} )$ &$2\cdot \one_{0}(T_3)$  &$-6$& $+6$ & $+6$ & $-6$ & $-12$ & $-9$ & $-3$ & $\frac{+124}{7}$ &  $+1  $& $+1 $ & $-1 $  \\
   $\sigma_{13}'$ & $(0^5;\frac{ 1}{2}\,\frac{1}{2}\,\frac{-1}{2}\,)(0^5;\frac{-1}{4}\,\frac{3}{4}\,\frac{-1}{2})'$&$\frac{+2}{3}$ &$2(1_1+1_3,1_{\bar{1}}+ 1_{\bar{3}} )$ &$ 2\cdot  \one_{0}(T_3)$  &$-6$& $+6$ & $+6$ & $-6$ & $-12$ & $-9$ & $-3$ & $\frac{+124}{7}$ &  $-1  $& $+1 $ & $+1 $  \\
  $\sigma_{14}$ & $(0^5;\frac{ 1}{2}\,\frac{1}{2}\,\frac{-1}{2}\,)(0^5;\frac{-1}{4}\,\frac{-1}{4}\,\frac{1}{2})'$&$\frac{+2}{3}$ &$2(1_{\bar{1}})+1( 1_{\bar{3}} )$ &$ 3\cdot  \one_{0}(T_3)$  &$+4$& $+6$ & $+6$ & $-6$ & $0$ & $+9$ & $+3$ & $\frac{+58}{7}$ &  $-1  $& $+1 $ & $+1 $  \\
$\sigma_{15}$  & $(0^5;\frac{-1}{2}\,\frac{-1}{2}\,\frac{-1}{2})(0^5;\frac{+3}{4}\,\frac{-1}{4}\,\frac{-1}{2})'$&$\frac{+2}{3}$ &$2(1_1+1_3,1_{\bar{1}}+ 1_{\bar{3}} )$& $ 2\cdot\one_{0}(T_3)$&$-6$& $-6$ & $-6$ & $-6$ & $+12$ & $-9$ & $-3$ & $\frac{-30}{7}$ &   $+1  $& $+1 $ & $-1 $  \\
$\sigma_{15}'$  & $(0^5;\frac{-1}{2}\,\frac{-1}{2}\,\frac{-1}{2})(0^5;\frac{+3}{4}\,\frac{-1}{4}\,\frac{-1}{2})'$&$0$ &$2(1_1+1_3,1_{\bar{1}}+ 1_{\bar{3}} )$& $  4\cdot \one_{0}(T_3)$  &$-6$& $-6$ & $-6$ & $-6$ & $+12$ & $-9$ & $-3$ & $\frac{-30}{7}$ &   $-1  $& $+1 $ & $+1 $  \\
  $\sigma_{16}$  & $(0^5;\frac{-1}{2}\,\frac{+1}{2}\,\frac{+1}{2})(0^5;\frac{+3}{4}\,\frac{-1}{4}\,\frac{-1}{2})'$&$\frac{+1}{3}$ & $2(1_1+1_3,1_{\bar{1}}+ 1_{\bar{3}} )$ &$ 2\cdot \one_{0}(T_3)$   &$-2$& $-6$ & $+6$ & $+6$ & $+12$ & $-9$ & $-3$ & $\frac{+40}{7}$ &  $+1  $& $+1 $ & $-1 $  \\
 $\sigma_{16}'$  & $(0^5;\frac{-1}{2}\,\frac{+1}{2}\,\frac{+1}{2})(0^5;\frac{+3}{4}\,\frac{-1}{4}\,\frac{-1}{2})'$&$\frac{+2}{3}$ & $2(1_1+1_3,1_{\bar{1}}+ 1_{\bar{3}} )$ &$ 2\cdot  \one_{0}(T_3)$   &$-2$& $-6$ & $+6$ & $+6$ & $+12$ & $-9$ & $-3$ & $\frac{+40}{7}$ &  $-1  $& $+1 $ & $+1 $  \\
$\sigma_{17}$ & $(0^5;\frac{+1}{2}\,\frac{+1}{2}\,\frac{-1}{2})(0^5;\frac{-1}{4}\,\frac{+3}{4}\,\frac{-1}{2})'$&$\frac{+1}{3}$ &$2(1_1+1_3,1_{\bar{1}}+ 1_{\bar{3}} )$ &$ 2\cdot \one_{0}(T_3)$   &$-6$& $+6$ & $+6$ & $-6$ & $-12$ & $-9$ & $-3$ & $\frac{+124}{7}$ &   $+1  $& $+1 $ & $-1 $  \\
$\sigma_{17}'$ & $(0^5;\frac{+1}{2}\,\frac{+1}{2}\,\frac{-1}{2})(0^5;\frac{-1}{4}\,\frac{+3}{4}\,\frac{-1}{2})'$&$\frac{+2}{3}$ &$2(1_1+1_3,1_{\bar{1}}+ 1_{\bar{3}} )$ &$ 2\cdot \one_{0}(T_3)$  &$-6$& $+6$ & $+6$ & $-6$ & $-12$ & $-9$ & $-3$ & $\frac{+124}{7}$ &   $-1  $& $+1 $ & $+1 $  \\
  $\sigma_{18}$  & $(0^5;\frac{ 1}{2}\,\frac{+1}{2}\,\frac{-1}{2})(0^5;\frac{+3}{4}\,\frac{-1}{4}\,\frac{-1}{2})'$&$\frac{+2}{3}$ & $2(1_{\bar{1}})+1( 1_{\bar{3}} )$ &$ 2\cdot  \one_{0}(T_3)$   &$+4$& $+6$ & $+6$ & $-6$ & $0$ & $+9$ & $+3$ & $\frac{+58}{7}$ &  $-1  $& $+1 $ & $+1 $  \\
$\sigma_{21}$  & $(0^5;\frac{-1}{6}\,\frac{-1}{6}\,\frac{-1}{6})(0^5;\frac{1}{4}\,\frac{1}{4}\,\frac{-1}{2})'$&$0$ &$1(1_{\bar1})$& $ \one_{0}(T_{1}^0)$ &$+2$& $-2$ & $-2$ & $-2$ & $0$ & $+9$ & $+3$ & $\frac{+12}{7}$ & $-1$& $-1$ & $-1$  \\ 
$\sigma_{22}$  & $(0^5;\frac{-5}{6}\,\frac{1}{6}\,\frac{1}{6})(0^5;\frac{1}{4}\,\frac{1}{4}\,\frac{-1}{2})'$&$0$ &$1(1_{\bar1}+1_3)$& $ \one_{0}(T_{5}^0)$ &$+2$& $-10$ & $+2$ & $+2$ & $0$ & $+9$ & $+3$ & $\frac{-2}{7}$ & $-1$& $+1$ & $+1$   \\
$\sigma_{23}$  & $(0^5;\frac{1}{6}\,\frac{-5}{6}\,\frac{1}{6})(0^5;\frac{1}{4}\,\frac{1}{4}\,\frac{-1}{2})'$&$0$ &$1(1_{\bar1}+1_3)$& $ \one_{0}(T_{5}^0)$ &$+2$& $-10$ & $+2$ & $+2$ & $0$ & $+9$ & $+3$ & $\frac{-44}{7}$ & $-1$& $+1$ & $+1$   \\
$\sigma_{24}$  & $(0^5;\frac{1}{6}\,\frac{1}{6}\,\frac{-5}{6})(0^5;\frac{1}{4}\,\frac{1}{4}\,\frac{-1}{2})'$&$0$ &$1(1_{\bar1}+1_3)$& $ \one_{0}(T_{5}^0)$ &$+2$& $-10$ & $+2$ & $+2$ & $0$ & $+9$ & $+3$ & $\frac{+82}{7}$ & $-1$& $+1$ & $+1$   \\[0.2em]
  \botrule
\end{tabular} 
\end{center}
\caption{U(1) charges of   L-handed neutral scalars (but $\sigma_{7,8}$ for R-handed).   We kept up to one oscillators represented as  {\it Number of resulting fields}$({\rm number~ of~ oscillating~ mode})$. For example, $n(1_{\bar{1}})$ means that there results $n$ multiplicities with one oscillator $\frac{-5}{12}$.  For $Q_{18,20,22}$ charges, here we listed only those of L-handed fields, participating in the Yukawa couplings.  $\sigma_{2,3,4,11',15',21,22,23,24}$ have phase $\Theta_i=0$, which can be used to break $\Z_{4R}$ down to $\Z_{2R}$.}\label{tab:Scalars} 
\end{table}

Since all the entries of the mass matrix $M^{(l)} $ are allowed, we show below how the required form (\ref{eq:Mweak1}) results. Because of the degeneracy of the SM fields in the sector $T_4^0$, the mass matrix can be written as
\dis{
\sim \begin{pmatrix} r_1 e^{i\phi_1}, &r_1 e^{i\phi_1}, &r_2 e^{i\phi_2}\\
 r_1 e^{i\phi_1}, &r_1 e^{i\phi_1}, &r_2 e^{i\phi_2}\\
  r_3 e^{i\phi_3}, &r_3 e^{i\phi_3}, &r_4 e^{i\phi_4}\\
\end{pmatrix}.
}
Redefining the L-handed and R-handed phases,
\dis{
l'_L=\begin{pmatrix} e^{i\phi_5}, &0, &0\\
 0, & e^{i\phi_5}, &0\\
   0 , &0, &1\\
\end{pmatrix}l_L,~l'_R=\begin{pmatrix} 1, &0, &0\\
 0, & 1, &0\\
   0 , &0, &e^{i\phi_6}\\
\end{pmatrix}l_R,
} 
we obtain the mass matrix for the choice of $\phi_5=-\phi_1$ and $\phi_6=\phi_2-\phi_1$,
\dis{
\sim \begin{pmatrix} r_1  , &r_1 , &r_2  \\
  r_1  , &r_1 , &r_2  \\
  r_3 e^{i\phi_3}, &r_3 e^{i\phi_3}, &r_4 e^{-i\phi_2}\\
\end{pmatrix}.
}
The above mass matrix form is simple enough to assign phases in the SM singlet fields, $\sigma_i$. From Eq. (\ref{eq:effMe1}), we can choose the following phase for the singlets, $\langle \sigma_1\rangle\sim e^{i\phi_3}, \langle \sigma_5\rangle\sim e^{i\cdot 0}$ and $\langle \sigma_2\rangle\sim e^{i\phi_2}$. Determining these phases  is postponed until a sufficiently accurate value of the PMNS phase is known.

\section{Conclusion}\label{sec:Conclusion} 

After presenting a useful parametrization in the Kim-Seo form of the PMNS matrix, we obtained the mass matrix forms of neutrinos and charged leptons from symmetries allowed in a compactified string \cite{Kim18Rp}.
The flipped SU(5) model compactified on  $\Z_{12-I}$ is simple enough to draw this analysis up to satisfying all data on the PMNS matrix. In the flipped SU(5), the $d$-type quark mass matrix and the neutrino mass matrix are symmetric. These matrices are set to be real. We have shown that the CP phase $\dell$ in the PMNS matrix can be introduced from the charged lepton mass matrix.

\acknowledgments{J.E.K. thanks Carlos Mu\~noz for the helpful discussion during his visit of UAM, Madrid, Spain, where this work was initiated. This work is supported in part by the IBS (IBS-R017-D1-2014-a00) and by the National Research Foundation (NRF) grant  NRF-2018R1A2A3074631.}
\newpage

\end{document}